\documentclass[aps,prb,twocolumn,showpacs,amssymb,superscriptaddress]{revtex4-1}

\usepackage{amsmath}  % needed for \tfrac, \bmatrix, etc.
\usepackage{amsfonts} % needed for bold Greek, Fraktur, and blackboard bold
\usepackage{graphicx} % needed for figures

\usepackage{color}			% for various colored (gray) things
\begin{document}

% Be sure to use the \title, \author, \affiliation, and \abstract macros
% to format your title page.  Don't use lower-level macros to  manually
% adjust the fonts and centering.

\title{What is superresolution microscopy?}

\author{John Bechhoefer}
\email{johnb@sfu.ca} 
\affiliation{Department of Physics, Simon Fraser University, Burnaby, BC V5A 1S6, Canada}

\date{\today}

\begin{abstract}
I explain what is, what is not, and what is only sort of superresolution microscopy.   I discuss optical resolution, first in terms of diffraction theory, then in terms of linear systems theory, and finally in terms of techniques that use prior information, nonlinearity, and other tricks to improve performance.  The discussion reveals two classes of superresolution:  Pseudo superresolution techniques improve images up to the diffraction limit but not much beyond.  True superresolution techniques allow substantial, useful improvements beyond the diffraction limit.  The two classes are distinguished by their scaling of resolution with photon counts.  Understanding the limits to imaging resolution involves concepts that pertain to almost any measurement problem, implying that the framework given here has broad application beyond optics.
\end{abstract}

\maketitle % title page is now complete

\section{Introduction} 
Until the 19th century, it was assumed that improving microscope images was a matter of reducing aberrations by grinding more accurate lenses and by using more sophisticated shapes in their design.  In the 1870s, Ernst Abbe\cite{volkmann66}  (with further contributions by Rayleigh\cite{rayleigh1896} in 1896 and Porter\cite{porter1906} in 1906) came to a radically different conclusion:  that wave optics and diffraction posed fundamental limits on the ability to image. These \textit{resolution limits} were  proportional to the wavelength $\lambda$ of light used and pertained to all wave-based imaging.  

Beginning in the 1950s, various researchers revisited the question of resolution limits, from the point of view of engineering and linear systems analysis.\cite{toraldo52,goodman05,huang09}  They noted that traditional discussions of diffraction limits ignored the intensity of images and argued that increasing brightness could, \textit{in principle}, increase resolution beyond the diffraction limit, a phenomenon they termed \textit{superresolution}.\cite{superres-engineers}  The words ``in principle" are key, because, \textit{in practice}, such techniques have never led to more than rudimentary demonstrations, although they have given important methods that improve the quality of imaging \textit{near} the diffraction limit.\cite{sibarita05} 

In the last 20 years, spectacular technological and conceptual advances have led to instruments that routinely surpass earlier diffraction limits, a phenomenon also termed ``superresolution."  Unlike the earlier work, these new techniques have led to numerous applications, particularly in biology,\cite{natmeth09,leung11} and commercial instruments have begun to appear.\cite{commercial}

Although the developments in the 1950s and in the last 20 years both concerned ``superresolution," the pace of recent advances makes it obvious that something has changed.  I will argue that there are two qualitatively different categories of superresolution techniques, one that gives ``pseudo" superresolution and another that leads to ``true" superresolution.  Sheppard\cite{sheppard07} and Mertz\cite{mertz10} have similarly classified superresolution methods; the somewhat different exposition here was inspired by an example from Harris's 1964 ``systems-style" discussion.\cite{harris64a}  

In the explosion of interest concerning superresolution techniques, the difference between these categories has sometimes been confused.  I hope this article will help clarify the situation.  Our discussion will focus on basic concepts rather than the details of specific schemes, for which there are excellent reviews.\cite{hell10,huang10}  A long, careful essay by Cremer and Masters gives a detailed history of superresolution and shares the view that key concepts have been re-invented or re-discovered many times.\cite{cremer13}

The discussion will be framed in terms of a simple imaging problem, that of distinguishing between one point source and two closely spaced ones.  In Sec.~\ref{sec:diff-limit}, we begin by reviewing the diffraction limit to optics and its role in limiting optical performance.  In Sec.~\ref{sec:linear-sys}, we discuss optical instrumentation from the point of view of linear-systems theory, where imaging is a kind of low-pass filter, with a resolution that depends on wavelength and signal strength (image brightness).  In Sec.~\ref{sec:priors}, we will consider the role of prior expectations in setting resolution.  It has long been known that special situations with additional prior information can greatly improve resolution; what is new is the ability to ``manufacture" prior expectations that then improve resolution, even when prior information would seem lacking.  In Sec.~\ref{sec:nonlinear}, we discuss how nonlinearity, by reducing the effective wavelength of light, is another approach to  surpassing the classical limits.  We will argue that these last two methods, prior engineering and nonlinearity, form a different, more powerful class of superresolution techniques than those based on linear-systems theory.  Finally, in Sec.~\ref{sec:conclusion}, we discuss some of the implications of our classification scheme.

\section{Resolution and the diffraction limit}
\label{sec:diff-limit}

The Abbe limit of resolution is textbook material in undergraduate optics courses.\cite{hecht02,brooker02,lipson11}  Based on an analysis of wave diffraction that includes the size of lenses and the imaging geometry, it gives the minimum distance $\Delta x$ that two objects can be distinguished:\cite{axial}
\begin{equation}
	\Delta x_{\rm Abbe} 
	= \frac{\lambda}{2n \, \sin \alpha} \equiv  \frac{\lambda}{2 \, \text{NA}} \,,
\label{eq:abbe-limit}
\end{equation}
where $n$ gives the index of refraction of the medium in which the imaging is done and where $\alpha$ is the maximum angle between the optical axis and all rays captured by the microscope objective.  NA $\equiv n \, \sin \alpha$ stands for \textit{numerical aperture} and is used to describe the resolution of microscope objectives.\cite{note:magnification}  A standard trick in microscopy is to image in oil, where $n \approx 1.5$.  The resolution improvement relative to air imaging is a factor of $n$ and corresponds to an effective wavelength $\lambda/n$ in the medium.  Well-designed objects can capture light nearly up to the maximum possible angle, $\alpha = \pi / 2$.  Thus, NA = 1.4 objectives are common and imply a resolution limit of $d \approx 180$ nm, at $\lambda = 500$ nm.  With proper sample preparation (to preclude aberrations), modern fluorescence microscopes routinely approach  this limit. 

To put the ideas of resolution in a more concrete setting, let us consider the problem of resolving two closely spaced point sources.  To simplify the analysis, we consider one-dimensional (1d) imaging with incoherent, monochromatic illumination.  Incoherence is typical in fluorescence microscopy, since each group emits independently, which implies that intensities add.  We also assume an imaging system with unit magnification.  Extensions to general optical systems, two dimensions, circular apertures, and coherent light are straightforward.  For perfectly coherent light, we would sum fields, rather than  intensities.  More generally, we could consider partially coherent light, using correlation functions. \cite{hecht02,brooker02,lipson11}

\begin{figure}[ht!]
	\centering
	\includegraphics[width=3.5in]{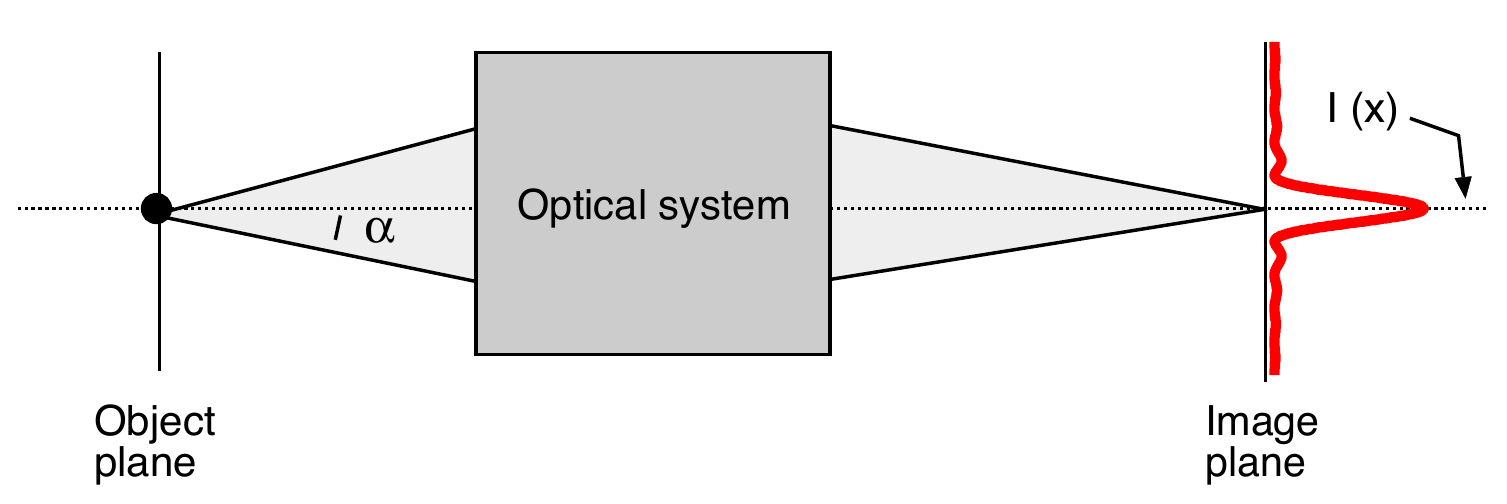}
	\caption{Schematic of imaging process, showing the point-spread function $I(x)$ in the image plane.  The maximum angle of rays collected, $\alpha$, determines the numerical aperture (NA).}
\label{fig:imagingSchematic}
\end{figure}

A standard textbook calculation\cite{hecht02,brooker02,lipson11,goodman05} shows that  the image of a point source $I^{(1)}_{\rm in}(x) = \delta(x)$ is the Fraunhofer diffraction pattern of the limiting aperture (exit pupil), which here is just a 1d slit.  The quantity $I(x)$ is the intensity, normalized to unity, of the image of a point object and is termed the \textit{point spread function} (PSF).  The function is here defined in the imaging plane (Fig.~\ref{fig:imagingSchematic}.)  Again, we consider a one-dimensional case where intensities vary in only one direction ($x$).

\begin{figure}[ht!]
	\centering
	\includegraphics[width=3.5in]{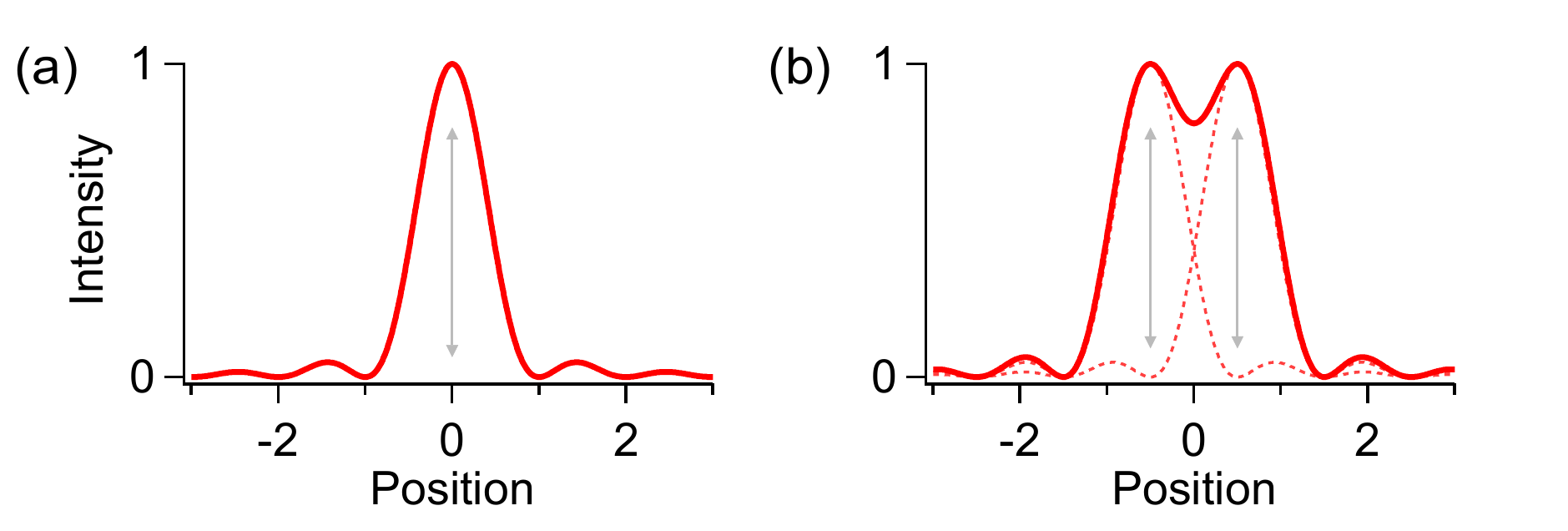}
	\caption{(a) Isolated point spread function.  (b)  Two point sources separated by $\Delta x_{\rm Abbe}$.}
\label{fig:2or1}
\end{figure}

Figure~\ref{fig:2or1}a shows the resulting point-spread function: 
\begin{equation}
	I^{(1)}_{\rm out}(x) 
		= \left[ \text{sinc} \left( \frac{2\pi \text{NA} \bar{x}}{\lambda} \right) \right]^2
		\equiv \text{sinc} (\pi x)^2  \,,
\label{eq:psf1}
\end{equation}
where $x = \bar{x}/(\Delta x_{\rm Abbe})$ is dimensionless.  We also consider the image formed by two point sources separated by $\Delta x$:
\begin{equation}
	I^{(2)}_{\rm out}(x) 
		= \text{sinc} \left[\pi \left( x-\tfrac{1}{2} \Delta x \right) \right]^2 
		+ \text{sinc} \left[\pi \left( x+\tfrac{1}{2} \Delta x \right) \right]^2 \,.
\label{eq:psf2}
\end{equation}
Figure~\ref{fig:2or1}b shows the image of two PSFs separated by $\Delta x = 1$ (or $\Delta \bar{x} = \Delta x_{\rm Abbe}$), illustrating the intensity profile expected at the classical diffraction limit.  The maximum of one PSF falls on the first zero of the second PSF, which also defines the Rayleigh resolution criterion, $\Delta x_{\rm Rayleigh}$.  (With circular lenses, the two criteria differ slightly.)  Traditionally, the Abbe/Rayleigh separation between sources defines the diffraction limit.  Of course, aberrations, defocusing, and other non-ideal imaging conditions can further degrade the resolution.   Below, we will explore techniques that allow one to infer details about objects at scales well below this Abbe/Rayleigh length.  

\section{Optics and linear systems}
\label{sec:linear-sys}

Much of optics operates in the \textit{linear response} regime, where the intensity of the image is proportional to the brightness of the source.  For an incoherent source, a general  optical image is the convolution between the ideal image of geometrical optics and the PSF:
\begin{align}
	I_{\rm out}(x) &= \int_{-\infty}^\infty dx \, G(x-x') \,I_{\rm in}(x') \,, \nonumber \\[3pt]
	 \tilde{I}_{\rm out}(k) &=  \tilde{G}(k) \, \tilde{I}_{\rm in}(k)  \,,
\label{eq:convolution}
\end{align}  
where the integration over $\pm \infty$ is truncated because image and object have finite extent.  The tilde indicates Fourier transform, defined as $\tilde{I}(k) = \int_{-\infty}^\infty dx \, e^{ikx} I(x)$ and $I(x) = \int_{-\infty}^\infty \tfrac{dk}{2\pi} e^{-ikx} \tilde{I}(k)$.  The second relation in Eq.~\eqref{eq:convolution} is just the convolution theorem.  The important physical point is that with incoherent illumination, intensities add---not fields.

This \textit{Fourier optics} view was developed by physicists and engineers in the mid-20th century, who sought to understand \textit{linear systems} in general.\cite{duffieux83,goodman05}  Lindberg gives a recent review.\cite{lindberg12}  One qualitatively new idea is to consider the effects of measurement noise, as quantified by the signal-to-noise ratio (SNR).  Let us assume that the intensity of light is set such that a detector, e.g., a pixel in a camera array, records an average of $N$ photons after integrating over a time $t$.   For high-enough light intensities, photon \textit{shot noise} usually dominates over other noise sources such as the electronic noise of charge amplifiers (\textit{read noise}), implying that if $N \gg 1$, the noise measured will be approximately Gaussian, with variance $\sigma^2 = N$.  

Since measuring an image yields a stochastic result, the problem of resolving two closely spaced objects can be viewed as a task of \textit{decision theory}:  given an image, did it come from one object or two?\cite{harris64a,small09,prasad14}  Of course, maybe it came from three, or four, or even more objects, but it will simplify matters to consider just two possibilities.  This statistical view of resolution will lead to criteria that depend on signal-to-noise ratios and thus differ from Rayleigh's ``geometrical" picture in terms of overlapping point-spread functions.

A systematic way to decide between scenarios is to calculate their likelihoods, in the sense of probability theory, and to choose the more likely one.  Will such a choice be correct?  Intuitively, it will if the difference between image models is much larger than the noise.  More formally, Harris (1964) calculates the logarithm of the ratio of likelihood functions.\cite{harris64a}  (Cf. the Appendix.)  We thus consider the SNR between the difference of image models and the noise:
\begin{align}
	\text{SNR} &= \frac{1}{\sigma^2} \int_{-\infty}^\infty dx \, 
	\left[ I^{(1)}_{\rm out}(x) -  I^{(2)}_{\rm out}(x) \right]^2 \nonumber \\[3pt]
	&= \frac{1}{\sigma^2} \int_{-\infty}^\infty \frac{dk}{2\pi} 
	\left| \tilde{I}^{(1)}_{\rm out}(k) -  \tilde{I}^{(2)}_{\rm out}(k) \right|^2 
		\nonumber\\[3pt]
	&= \frac{1}{\sigma^2} \int_{-\infty}^\infty \frac{dk}{2\pi} \, 
	\left| \tilde{I}^{(1)}_{\rm in}(k) -  \tilde{I}^{(2)}_{\rm in}(k) \right|^2 \,
	\left| \tilde{G}(k) \right|^2 \,,
\label{eq:snr}
\end{align}
where we use Parseval's Theorem in the second line and Eq.~\eqref{eq:convolution} in the third.  The $\sigma^2$ factor represents the noise---the variance per length of photon counts for a measurement lasting a time $t$.

The Fourier transforms of the input image models are given by $\tilde{I}^{(1)}_{\rm in}(k) = 1$ and
\begin{align}
	\tilde{I}^{(2)}_{\rm in}(k) &= \int_{-\infty}^\infty dx \, \tfrac{1}{2} \left[ 
	\delta \left( x-\tfrac{1}{2} \Delta x \right) 
		+ \delta \left( x+\tfrac{1}{2} \Delta x \right) \right] \, e^{ikx}  \nonumber \\
		&= \cos \left( \tfrac{1}{2}k\Delta x \right) \,.
\label{eq:ft}
\end{align}
To calculate the signal-to-noise ratio, we note that intensities are proportional to the photon flux and the integration time $t$.  Since shot noise is a Poisson process, the variance $\sigma^2 \sim t$.  By contrast, for the intensities, $I^2 \sim t^2$, and the SNR is thus proportional to $t^2/t = t$.  Using incoherent light implies that $G(x)$ is the intensity response and hence that $\tilde{G}(k)$ is the autocorrelation function of the pupil's transmission function.\cite{goodman05}  For a 1d slit, $\tilde{G}(k)$ is the triangle function, equal to $1-|k|/k_{\rm max}$ for $|k| < k_{\rm max}$ and zero for higher wavenumbers.\cite{goodman05} The cutoff frequency is $k_{\rm max} = 2\pi/\Delta x_{\rm Abbe}$.  Including the time scaling, Eq.~\eqref{eq:snr} then becomes 
\begin{equation}
	\text{SNR} \propto t \int_{-k_{\rm max}}^{k_{\rm max}} dk \, 
	\left[ 1 -  \cos \left( \tfrac{1}{2}k\Delta x \right) \right]^2 \, (1-|k|/k_{\rm max})^2\,.
\label{eq:snr2}
\end{equation}

To compute the SNR for small $\Delta x$, consider the limit $k_{\rm max} \Delta x \ll 1$  and expand the integrand as $\bigl[ 1-[1-\tfrac{1}{2} (\tfrac{1}{2} k \Delta x)^2 + \cdots] \bigr]^2 (1-\cdots)^2 \approx \bigl[ \tfrac{1}{8} (k \Delta x)^2 \bigr]^2$.  Thus, the SNR $\sim t \, (k_{\rm max} \Delta x)^4$, or
\begin{equation}
	\Delta x \sim \Delta x_{\rm Abbe} \, N^{-1/4} \,,
\label{eq:snr3}
\end{equation}
where we replace time with the number of photons detected $N$ and assume that detection requires a minimum value of SNR, kept constant as $N$ varies.   A modest increase in resolution requires a large increase in photon number.  The  unfavorable scaling explains why the strategy of increasing spatial resolution by  boosting spatial frequencies beyond the cutoff cannot  increase resolution more than marginally:  the signal disappears too quickly as $\Delta \bar{x}$ is decreased below $\Delta x_{\rm Abbe}$.  

\begin{figure}[t]
	\centering
	\includegraphics[width=2.5in]{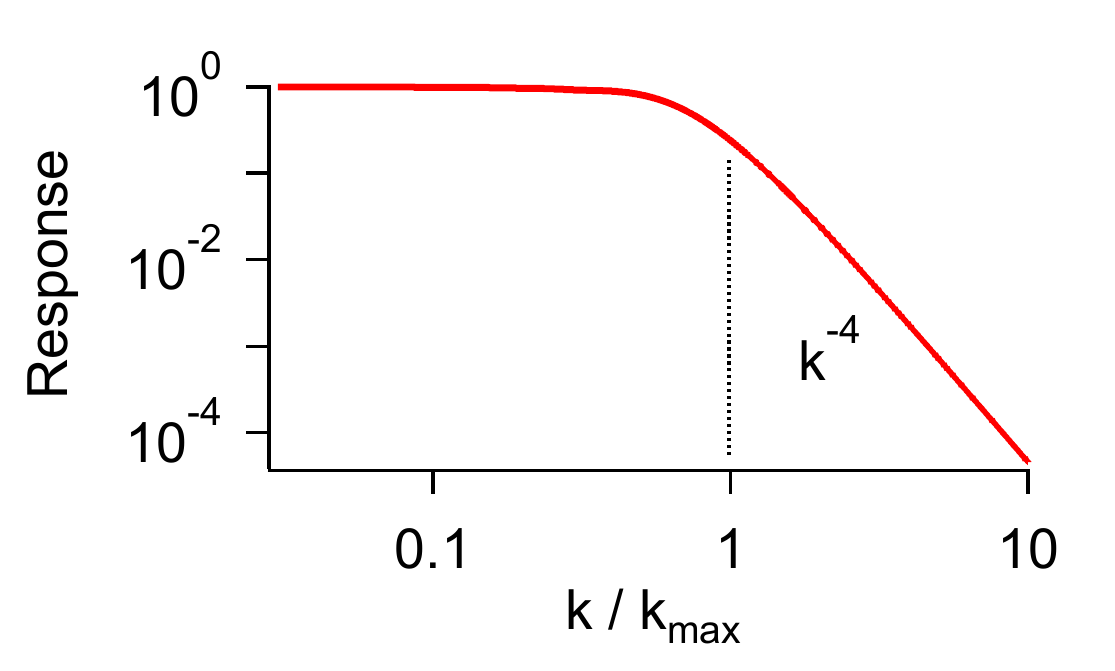}
	\caption{Modulation transfer function vs. wavenumber, with cutoff frequency $k/k_{\rm max} \equiv  \Delta x_{\rm Abbe} / \Delta \bar{x} = 1$ (vertical dotted line). }
\label{fig:bode}
\end{figure}

Returning from the small-$\Delta x$ limit summarized by Eq.~\eqref{eq:snr3} to the full expression for SNR, Eq.~\eqref{eq:snr2}, is plotted as Fig.~\ref{fig:bode}, which is normalized to have unity gain for large $\Delta x$.  We see that the amplitude transfer function for the difference model has the form of a low-pass filter.\cite{low-pass}  Spatial frequencies below the cutoff are imaged faithfully, but information is severely attenuated when $k> k_{\rm max}$.  

Although the $N^{-1/4}$ scaling law is supported by the analysis of a specific case, the exponent is generic.  Essentially, we distinguish between two possible intensity profiles that have different widths, or, equivalently, between two probability distributions that have different variances.  The $-1/4$ exponent in the $N^{-1/4}$  scaling law then reflects a ``variance of variance."\cite{lucy92}

If boosting attenuated signals does not lead to significant resolution gains, it can still be very effective in ``cleaning up" images and allowing them to approach the standard diffraction limit.  Indeed, signal-processing techniques lead to \textit{deconvolution microscopy}, which is a powerful approach to image processing that, with increasing computer power, is now quite practical.\cite{sibarita05} But attempts to use similar techniques to exceed the diffraction limit\cite{piche12}---what I call pseudo superresolution---can have only very limited success.  The same conclusion pertains to ``hardware strategies" that try to modify, or ``engineer" the pupil aperture function to reduce the spot size.\cite{toraldo52,ramsay08}  

A more general way to understand some of the limitations of these classical superresolution approaches is to use \textit{information theory}.\cite{toraldo55,lipson03}  One insight that information theory provides is that an optical system has a finite number of \textit{degrees of freedom}, which is proportional to the product of spatial and temporal bandwidths.  The number of degrees of freedom is fixed in an optical system, but one can trade off factors.  Thus, one can increase spatial resolution at the expense of temporal resolution.  This is another way of understanding why collecting more photons can increase resolution.\cite{lukosz66, lukosz67,sheppard07,lindberg12}   However, it is too soon to give the last word on ways to understand resolution, as the spectacular advances in microscopy discussed in this article are suggesting new ideas and statistical tools that try, for example, to generalize measures of localization to cases where objects are labeled very densely by fluorophores.\cite{mukamel12,fitzgerald12,nieuwenhuizen13}

\section{Superresolution from ``prior engineering"}
\label{sec:priors}

In the last two decades, conceptual and practical breakthroughs have led to ``true" superresolution imaging, where the amount of information that can be recovered from an image by equivalent numbers of photons is greatly increased relative to what is possible in deconvolution microscopy.  In this section, we discuss an approach that depends on the manipulation, or``engineering," of prior knowledge.

\subsubsection{Reconstruction using prior knowledge can exceed the Abbe limit}

Abbe's diffraction limit implicitly assumed that there is no significant prior information available about the object being imaged.  When there is, the increase in precision of measurements can be spectacular.  As a basic example, we consider the \textit{localization} of a single source that we know to be isolated.  Here,  ``localization"  contrasts with ``resolution," which pertains to non-isolated sources.  This prior knowledge that the source is isolated makes all the difference.  If we think of our measurement ``photon by photon," the point-spread function becomes a unimodal probability distribution whose standard deviation $\sigma_0$ is set by the Abbe diffraction limit.  If we record $N$ independent photons, then the average has a standard deviation $\approx\sigma_0/\sqrt{N}$, as dictated by the Central Limit Theorem.\cite{sivia06} Thus, localization improves with increasing photon counts.\cite{bobroff86,ober04,mortensen10,deschout14}   For well-chosen synthetic fluorophores, one can detect $\mathcal{O}(10^4)$ photons, implying localization on the order of a nanometer.\cite{yildiz05}  (In live-cell imaging using fluorescent proteins, performance is somewhat worse, as only 100--2000 photons per fluorophore are typically detectable.\cite{patterson10})  Again:  localization is not the same as resolution, as it depends on prior information about the source.

\subsubsection{Reconstruction without prior knowledge fails}

\begin{figure}[ht!]
	\centering
	\includegraphics[width=3.5in]{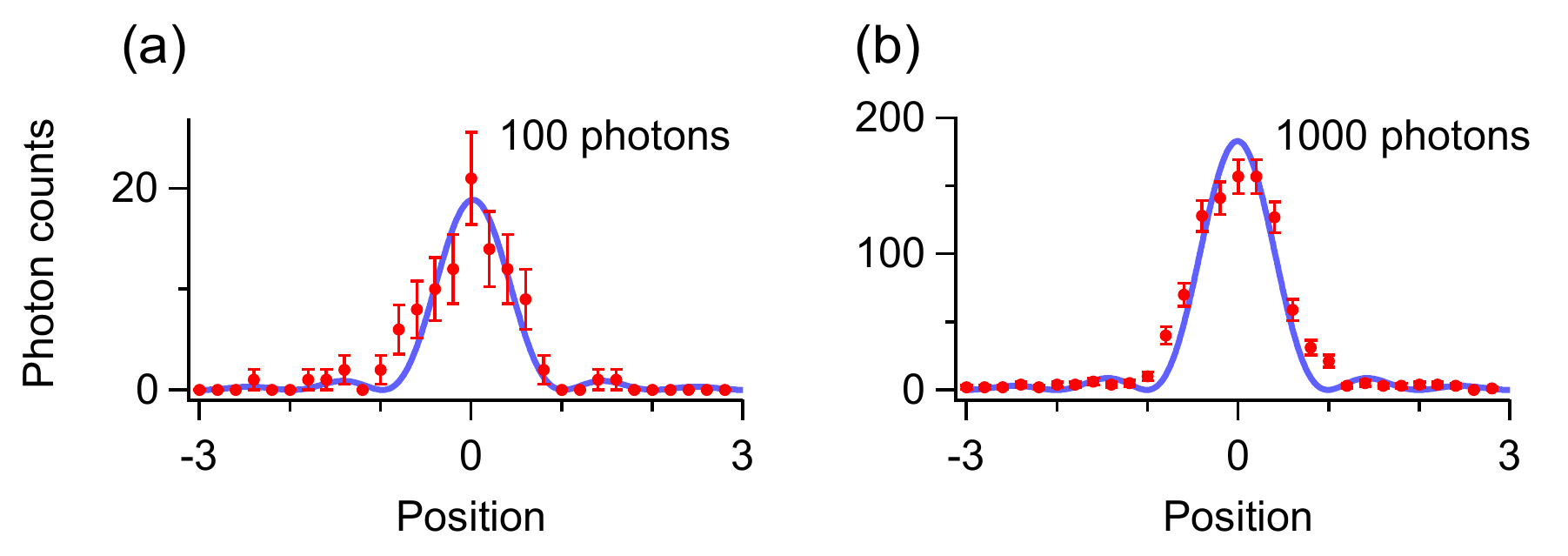}
	\caption{Two sources or one? (a) and (b) simulate two sources located at  $x=\pm 0.25$.  Each PSF has width = 1.  Markers show photon counts $N_i$ in each bin (pixel), with error bars equal to $\sqrt{N_i}$.  (a)  $100$ photons.  $\chi^2 = 14.2$ for $\nu = 14$ degrees of freedom. (b) $1000$ photons.  $\chi^2 = 120$ for $\nu = 27$.  The range $[-3,3]$ is divided into 30 bins.}
\label{fig:fewSource}
\end{figure}

We contrast the success in localizing a fluorophore that is known to be isolated with the failure that occurs when we do not know whether the fluorophore is isolated or not.   In Sec.~\ref{sec:linear-sys}, we considered the problem of distinguishing two sources from one and gave a scaling argument that for separations $\Delta \bar{x} \ll \Delta x_{\rm Abbe}$, the number of photons needed to decide between the two scenarios grows too rapidly to be useful.  Here, we show more intuitively that the task is hopeless.  In Fig.~\ref{fig:fewSource}, we simulate images from two point sources (Eq.~\eqref{eq:psf2}) separated by $\Delta \bar{x} = \tfrac{1}{2} \Delta x_{\rm Abbe}$.  The markers show the number of photon counts for each spatial bin (camera pixel), assuming measurements are shot noise limited.  Error bars are estimated as the square root of the number of counts in this Poisson process.\cite{errorbars}    In Fig.~\ref{fig:fewSource}a,  there are $\approx 100$ photon counts recorded.  A fit to a \textit{single} source, of unknown position and strength and width fixed to that of the PSF has a $\chi^2$ statistic that cannot be ruled out as unlikely.  The only way to distinguish between two sources and a single source would be to compare its amplitude to that of a single source, but sources can have different strengths:  Different types of fluorophores obviously do, but even a single type of fluorophore can vary in brightness.  For example, when immobilized on a surface and illuminated by polarized light, a molecule with fixed dipole moment emits photons at varying rates, depending on its orientation.\cite{betzig93,ha99,engelhardt11}  More fundamentally, all known types of fluorophores blink (emit intermittently \cite{frantsuzov08}), meaning that two measurements over long times of the integrated intensity of the same molecule can differ by amounts that greatly exceed the statistical fluctuations of a constant-rate emitter.

Increasing the counts to $\approx 1000$ (Fig.~\ref{fig:fewSource}b) allows one to rule out a single, constant-emitter-rate source, as the width now exceeds that of the PSF by a statistically significant amount.  (Note the smaller error bars for each point.)  Still, the disagreement is subtle, at best:  Reliable inference is unlikely without sufficient prior information.

\subsubsection{Stochastic localization: engineering the prior}  

Recently, two groups independently developed a technique that gives the precision of single-source localization microscopy without the need for \textit{a priori} knowledge of localization.  One version is known as PALM (Photo-Activated Localization Microscopy\cite{betzig06}) and another as STORM (Stochastic Optical Reconstruction Microscopy\cite{rust06}), and we will refer to them collectively as \textit{stochastic localization}.  They share the idea of making nearby molecules \textit{different}, using some kind of stochastic activation process, so that they can be \textit{separately} localized.\cite{precursors-stochastic}  One way to differentiate neighboring fluorophores is that some types of fluorescent groups are dark until photo-activated, usually by blue or UV light.\cite{patterson10}  Once active, the molecules may be excited fluorescently using lower-wavelength light.  Once excited, they fluoresce at a still-lower wavelength.  Thus, stochastic localization proceeds as follows:  A weak light pulse activates a random, \textit{sparse subset} of fluorophore molecules.  Each of these now-separated sources is then localized, as for isolated molecules.  After localization, the molecules should become dark again.  A simple way of ensuring this is to use a strong excitation pulse that photobleaches the active molecules, making them permanently dark.  Another activation pulse then turns on a different sparse subset, which is subsequently localized.  Repeating this cycle many times builds up an image whose sources are very close to each other.  The trick is to sequentially activate the sources, so that they are isolated while being interrogated.\cite{sparse}  We make sure that it is unlikely for more than one molecule to be activated in an area set by the diffraction length.  This knowledge functions as a kind of prior information.  In practice, it is not necessary to permanently photobleach molecules: one can take advantage of almost any kind of switching between active and dark states,\cite{dertinger09} as well as other kinds of prior information.\cite{berro12}

Thus, clever ``engineering" of prior expectations can give the benefits of localization microscopy, even when sources are not well-separated.  The precision is increased by $\sqrt{N}$ over the classical diffraction limit, where $N$ is the average number of photons recorded from a point source in one camera frame.

\section{Superresolution from nonlinearity}
\label{sec:nonlinear}

While stochastic localization is computationally based, an alternate technique known as STED (STimulated Emission Depletion) microscopy is ``hardware based." The idea was proposed in 1994 by Hell and Wichmann\cite{hell94} and then extensively developed in the former's group, along with a set of closely related methods.\cite{hell10}

\begin{figure}[ht!]
	\centering
	\includegraphics[width=3.0in]{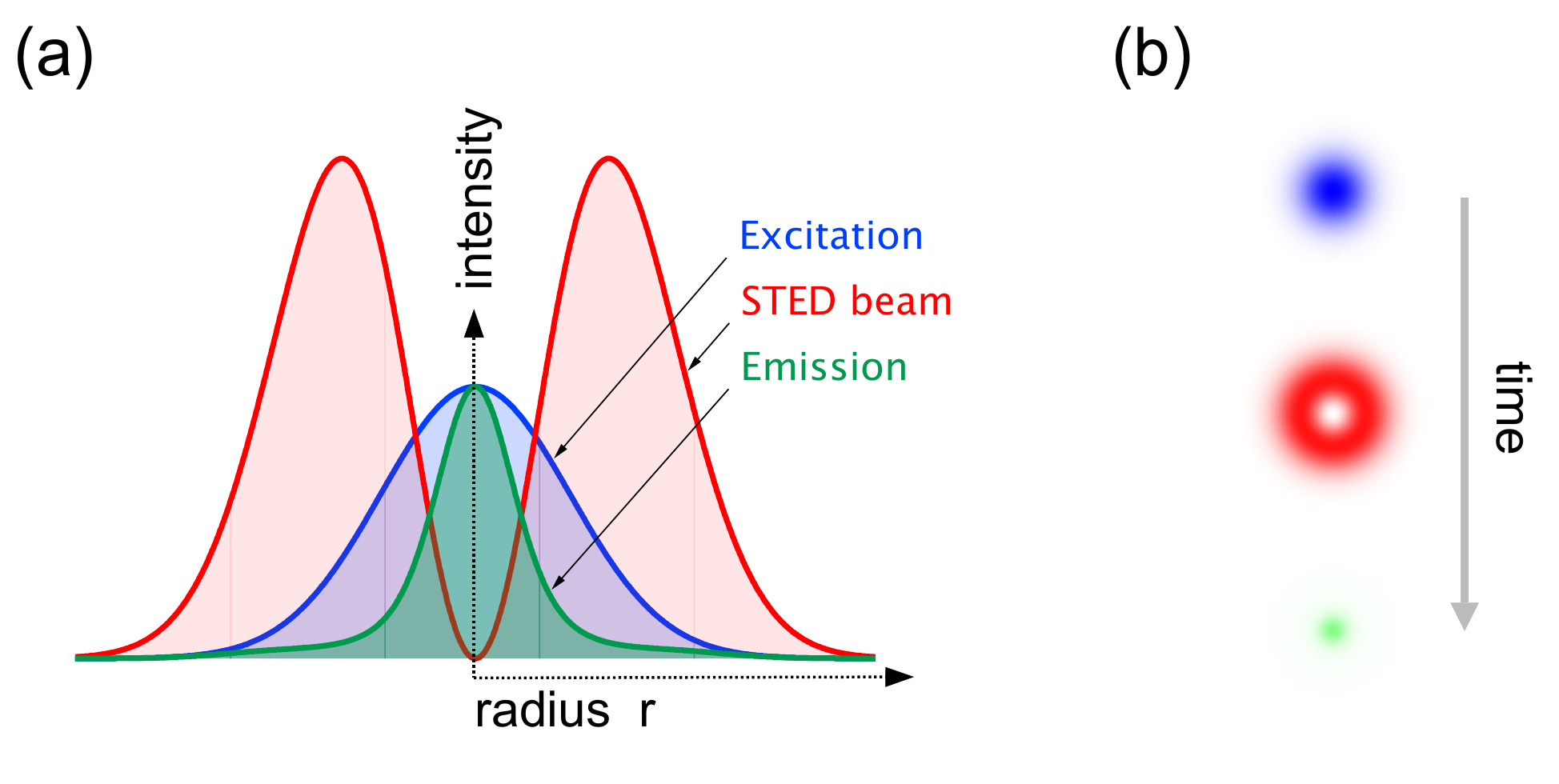}
	\caption{Illustration of STED imaging.  (a)  1d cut through intensity profile, illustrating the broad excitation pulse (blue), the doughnut-shaped STED depletion beam (red), and the narrow emission pulse (green).  (b)  2d beam profiles showing the temporal sequence of beams.}
\label{fig:STED}
\end{figure}

The basic idea of STED is illustrated in Fig.~\ref{fig:STED}.  A picosecond (ps)-scale conventional focussed spot excites fluorescence in a spot (blue).  The width of this beam (in the sample plane) has a scale set by the Abbe limit, $\lambda/(2\text{NA})$.  The excitation beam is followed by the ps-scale STED beam (red) a few ps after the original excitation pulse.  The timing ensures that the excited fluorescent molecules have not  had time to decay.  (Their lifetime $\approx$ ns.)  Because the STED beam has a dark spot at its center, it de-excites the original beam  ``from the outside in," using stimulated emission.  The distribution of surviving excited molecules then has a reduced width.  When they eventually decay, they are detected by their ordinary fluorescence emission (green).  The result is equivalent to a narrower excitation beam.  The reduced size of the point-spread function implies higher resolution.

The width of the emission point-spread function is given by\cite{harke08}
\begin{equation}
	\Delta x_{\rm STED} = \frac{\Delta x_{\rm Abbe}}{\sqrt{1+ I^{(0)}_{\rm STED}/I_{\rm sat}}} \,,
\label{eq:STEDresolution}
\end{equation}
where $I^{(0)}_{\rm STED}$ is the intensity scale of the de-excitation beam and where $I_{\rm sat}$ is the intensity at which the rate of absorption by the ground state matches the rate of emission by the excited state.  Physically, it depends on the cross section for stimulated emission.\cite{dyba05}  For $I^{(0)}_{\rm STED} \gg I_{\rm sat}$, $\Delta x_{\rm STED} \sim [I^{(0)}_{\rm STED}]^{-1/2} \sim \Delta x_{\rm Abbe} \, N^{-1/2}$, where $N$ is the number of photons in the STED beam.  The resolution improvement has the same scaling with photon counts as have stochastic localization techniques (indeed, localization in general).  Both are qualitatively better than the scaling for deconvolution microscopy.

We derive Eq.~\eqref{eq:STEDresolution} following Harke et al.\cite{harke08}  The 1d-excitation point-spread function in the sample plane is approximately $h_{\rm exc}(x) \sim e^{-x^2/2}$, with $x$ again in units of $\Delta x_{\rm Abbe}$.  We can approximate the STED beam intensity near the center by its quadratic expansion, so that  $I_{\rm STED}(x) \sim \tfrac{1}{2} [I^{(0)}_{\rm STED} / I_{\rm sat}] \, x^2$.  The constant factor, $I^{(0)}_{\rm STED} / I_{\rm sat}$, is the de-excitation beam intensity scale $I^{(0)}_{\rm STED}$, in units of $I_{\rm sat}$.  The STED pulse is approximated as a simple, constant-rate relaxation so that, as in a Poisson process, the fraction of surviving molecules in the original pulse is $\eta(x) \sim e^{-\tfrac{1}{2}(I^{(0)}_{\rm STED} / I_{\rm sat}) \, x^2}$.  (The same type of law holds for radioactive decay, with $\eta$ in that case being the fraction of molecules that survive after a given time.  In this interpretation, $I_{\rm sat}$ is analogous to a 1/$e$ lifetime at $x=1$.)  Thus,
\begin{align}
	h(x) &\sim h_{\rm exc}(x) \, \eta(x) \sim e^{-\tfrac{1}{2} [1+ (I^{(0)}_{\rm STED} / I_{\rm sat})] \, x^2} \nonumber \\
	&\equiv e^{-\tfrac{1}{2} (x^2/\Delta x_{\rm STED})^2} \,,
\label{eq:STEDemission}
\end{align}
which leads directly to Eq.~\eqref{eq:STEDresolution}.

Why is there a fundamental improvement in resolution?  STED is a nonlinear technique, and nonlinearity can improve the resolution by ``sharpening" responses.  For example, a response $\sim I(x)^2$ transforms a Gaussian point-spread function from $I(x) \sim \exp(-x^2/2\sigma^2)$ to $I(x)^2 \sim \exp(-x^2/\sigma^2)$, which has a width that is smaller by $\sqrt{2}$.  In STED, the key nonlinearity occurs in the exponential survival probability $\eta(x)$.  With a purely linear response, no resolution enhancement would be possible, since the spatial scale of the STED beam is also subject to the Abbe limit and must thus vary on the same length scale as the original excitation beam.

Stochastic localization and STED are just two among many techniques for fundamentally surpassing the classical diffraction limit.  For want of space, we omit  discussion of many other ways to surpass the Abbe limit, including pseudo superresolution techniques such as confocal imaging,\cite{pawley06} multiphoton microscopy,\cite{diaspro05}  and 4Pi-microscopy \cite{cremer78,hell92}; true superresolution techniques such as near-field scanning (NSOM),\cite{synge28,betzig86,novotny12} multiple scattering (which converts evanescent modes into propagating ones),\cite{simonetti06} saturation microscopy,\cite{heintzmann02,fujita07} and the ``perfect imaging" promised by metamaterials.\cite{pendry00,fang05}  Some techniques, such as structured illumination,\cite{gustafsson00,gustafsson05} are hard to classify because they contain elements of both types of superresolution.  Finally, although our discussion has focused on what is possible with classical light sources, we note that $N$ entangled nonclassical photon-number states can create interference patterns with wavelength $\lambda/2N$,\cite{boto00} an idea that has been partly implemented using a 4-photon state.\cite{rozema14}  Unfortunately, the efficiency of all quantum-optics schemes implemented to date is well below that of the classical methods we have been discussing.  Still, although practical applications seem far off, using light in $N$-photon entangled states promises imaging whose resolution can improve as $N^{-1}$.

\section{Conclusion and implications}
\label{sec:conclusion}

Superresolution microscopy techniques divide into two broad classes:
\begin{itemize}
\item Pseudo superresolution, based on deconvolution microscopy and other ideas of linear systems theory, which aims to make maximum use of the available information, using minimal prior expectations.  The general idea is to use the known, or estimated optical transfer function to boost the measured signal back to its original level.  The ability to do so is limited by measurement noise.  The poor scaling, $\Delta x \sim N^{-1/4}$, implies a resolution only slightly beyond the standard diffraction limit.

\item True superresolution, which increases the amount of recoverable information, for example by creating prior information (stochastic localization methods) or nonlinear tricks, such as those used in STED.  Resolution scales as $\Delta x \sim N^{-1/2}$, a much more favorable law that allows significant increases in resolution, in practical situations.  Potentially, light using nonclassical photon states can improve the scaling further, a situation we include in the category of true superresolution.
\end{itemize}

The classification of superresolution presented here is general and applies beyond  optics.  To list just one example, there is good evidence\cite{oppenheim13} that humans can resolve musical pitch much better than the classic time-frequency uncertainty principle, which states that the product $\Delta t \, \Delta f \ge \tfrac{1}{4\pi}$, where $\Delta t$ is the time a note is played and $\Delta f$ the difference in pitch to be distinguished.  Since humans can routinely beat this limit, Oppenheim and Magnasco conclude that the ear and/or  brain must use nonlinear processing.\cite{oppenheim13}  But louder sounds will also improve pitch resolution, in analogy with our discussion of light intensity and low-pass filtering, an effect they do not discuss.  Whether  ``audio superresolution" is due to high signal levels or to nonlinear processing, the ideas presented are perhaps useful for understanding the limits to pitch resolution.  

The questions about superresolution that we have explored here in the context of microscopy (and, briefly, human hearing) apply in some sense to any measurement problem.  Thus,  understanding what limits measurements---appreciating the roles of signal-to-noise ratio and of prior expectations---should be part of the education of a physicist.

\begin{acknowledgments}
I thank Jari Lindberg and Jeff Salvail for a careful reading of the manuscript and for valuable suggestions.
\end{acknowledgments}

\appendix*
\section{Decision making and the signal-to-noise ratio}

To justify more carefully the link between likelihood and signal-to-noise ratios, we follow Harris\cite{harris64a} and consider the problem of deciding whether a given image comes from Object 1 or Object 2.  (See Fig.~\ref{fig:2or1}.)  If the measured intensity were noiseless, the one-dimensional image would be either $I_{\rm out}^{(1)}(x)$ or $I_{\rm out}^{(2)}(x)$.  Let the image have pixels indexed by $i$ that are centered on $x_i$, of width $\Delta x$.  Let the measured intensity at each pixel be $I_i$.  The noise variance in one pixel $\sigma_p^2$ is due to shot noise, read noise, and dark noise, and its distribution is assumed Gaussian and independent of $i$, for simplicity.  (If the intensity varies considerably over the image, then we can define a $\sigma_p$ that represents an  average noise level.)   The likelihood that the image comes from Object 1 is then 
\begin{equation}
	L^{(1)} \approx \prod_i \frac{1}{\sqrt{2\pi}\sigma_p} \, 
		e^{-\frac{1}{2\sigma_p^2} \left[ I_i - I^{(1)}_i \right]^2} \,,
\end{equation}
where  $I^{(1)}_i \equiv I_{\rm out}^{(1)}(x)|_{x=x_i} \, \Delta x$ is the number of photons detected in pixel $i$ and the product is over all pixels in the detector.  An analogous expression holds for $L^{(2)}$.  Then the natural logarithm of the likelihood ratio is given by
\begin{equation}
	\psi_{12} \equiv \ln \frac{L^{(1)}}{L^{(2)}} = \frac{1}{2\sigma_p^2} \, 
	\sum_i \left\{ [I_i - I^{(2)}_i]^2 - [I_i - I^{(1)}_i]^2 \right\} \,.
\label{eq:likelihood-ratio}
\end{equation}
If Object 1 actually produces the image, then $I_i = I^{(1)}_i + n_i$, and Eq.~\eqref{eq:likelihood-ratio} becomes
\begin{equation}
	\psi_{12} = \sum_i  \left\{ \frac{1}{2\sigma_p^2} \left[ I^{(1)}_i-I^{(2)}_i \right]^2 
		- \frac{2 n_i}{2 \sigma_p^2} \left[ I^{(1)}_i-I^{(2)}_i \right] \right\} \,.
\end{equation}
If $n_i$ is Gaussian, so is $\psi_{12}$.  Its  mean is given by $\langle \psi \rangle = \tfrac{1}{2\sigma_p^2} \sum_i  [ I^{(1)}_i-I^{(2)}_i ]^2$ and its variance by $\sigma_\psi^2 =  \tfrac{1}{\sigma_p^2}  \sum_i  [ I^{(1)}_i-I^{(2)}_i ]^2$.  We will conclude that Object 1 produced the image if the random variable $\psi_{12} > 0$.  The probability that our decision is correct is thus given by
\begin{align}
	P(\psi_{12} > 0) &= \frac{1}{\sqrt{2\pi}\sigma_\psi} 
	\int_0^\infty d\psi \, e^{-\frac{(\psi-\langle \psi \rangle)^2}{2\sigma_\psi^2}} \nonumber \\
		&= \frac{1}{\sqrt{2\pi}} \int_{-\tfrac{\langle \psi \rangle}{\sigma_\psi}}^\infty dz \, e^{-\frac{z^2}{2}}  \nonumber \\
		&= \tfrac{1}{2} \left[1+\text{erf}{\sqrt{\text{SNR}}} \right]\,,
\end{align}
which depends only on $\tfrac{2\langle \psi \rangle}{\sigma_\psi} \equiv \sqrt{\text{SNR}}$.  Below, we show SNR to be the signal-to-noise ratio.  For SNR $\gg 1$, the probability of a \textit{wrong} decision is 
\begin{equation}
	1-P(\psi_{12} > 0) \sim \frac{1}{\sqrt{4\pi \, \text{SNR}}} \, e^{-\text{SNR}} \,,
\end{equation}
which rapidly goes to zero for large SNR.  To further interpret the SNR, we write
\begin{align}
	\text{SNR} &= \left( \frac{2 \langle \psi \rangle}{\sigma_\psi} \right)^2  
	= \frac{4\sigma_p^2}{4\sigma_p^4}\sum_i \left\{ [ I^{(1)}_i-I^{(2)}_i ]^2 \right\} \nonumber \\
	&\approx  \frac{\Delta x}{\sigma_p^2} 
		\int_{-\infty}^\infty dx \, \left\{ [ I_{\rm out}^{(1)}(x)-I_{\rm out}^{(2)}(x) ]^2 \right\} \,.
\label{eq:snr4}
\end{align}
Defining $\sigma^2 = \sigma_p^2/(\Delta x)$ to be the photon variance per length gives Eq.~\eqref{eq:snr}.  Recalling that $I_{\rm out}(x)$ is the number of photons detected per length in the absence of noise, we verify that the right-hand side of Eq.~\eqref{eq:snr4} is dimensionless.  Thus, $\sqrt{\text{SNR}}$ is the ratio of the photon count difference to the photon count fluctuation over  a given length of the image.


\begin{thebibliography}{99}
% The numeral (here 99) in curly braces is nominally the number of entries in
% the bibliography. It's supposed to affect the amount of space around the
% numerical labels, so only the number of digits should matter--and even that
% seems to make no discernible difference.

\bibitem{volkmann66}  H. Volkmann, ``Ernst {A}bbe and his work," Appl. Opt. \textbf{5}, 1720--1731 (1966).

\bibitem{rayleigh1896} L. Rayleigh, ``On the theory of optical images, with special reference to the microscope," The London, Edinburgh, and Dublin Phil. Mag. and Journal of Science \textbf{42}, Part XV, 167--195 (1896).

\bibitem{porter1906} A. B. Porter, ``On the diffraction theory of microscopic vision," The London, Edinburgh, and Dublin Phil. Mag. and Journal of Science \textbf{11}, 154--166 (1906).

\bibitem{toraldo52} G. Toraldo di Francia, ``Super-gain antennas and optical resolving power," Nuovo Cimento Suppl. \textbf{9}, 426--438 (1952).

\bibitem{goodman05} Joseph W. Goodman, \textit{Introduction to Fourier Optics}, 3rd ed. (Roberts and Company Publishers, 2005).  The first edition was published in  1968.

\bibitem{huang09} Fu Min Huang and Nikolay I. Zheludev, ``Super-resolution without evanescent waves," Nano Lett. \textbf{9}, 1249--1254 (2009).  The authors give a modern implementation of the aperture schemes pioneered by Toraldo Di Francia.\cite{toraldo52}

\bibitem{superres-engineers}  ``Superresolution" is also sometimes used to describe sub-pixel resolution in an imaging detector.  Since pixels are not necessarily related to intrinsic resolution, we do not consider such techniques here. 

\bibitem{sibarita05}  Jean-Baptiste Sibarita, ``Deconvolution microscopy," Adv. Biochem. Engin. / Biotechnol. \textbf{95}, 201--243 (2005).

\bibitem{natmeth09}  Superresolution fluorescence microscopy was the 2008 ``Method of the Year" for \textit{Nature Methods}, and its January 2009 issue contains commentary and interviews with scientists playing a principal role in its development.  This is a good ``cultural" reference.

\bibitem{leung11}  Bonnie O. Leung and Keng C. Chou, ``Review of superresolution fluorescence microscopy for biology," Appl. Spect. \textbf{65}, 967--980 (2011).

\bibitem{commercial} For example, a STED microscope is sold by the Leica Corporation.

\bibitem{sheppard07} Colin J. R. Sheppard, ``Fundamentals of superresolution," Micron \textbf{38}, 165--169 (2007).  Sheppard introduces three classes rather than two:   \textit{Improved} superresolution boosts spatial frequency response but leaves the  cutoff frequency unchanged.  \textit{Restricted} superresolution includes tricks that increase the cut-off by up to a factor of two.  We use ``pseudo" superresolution for both cases.  Finally, \textit{unrestricted} superresolution refers to what we term ``true" superresolution.

\bibitem{mertz10} Jerome Mertz, \textit{Introduction to Optical Microscopy}, (Roberts and Co., 2010), Ch. 18.  Mertz follows Sheppard's classification, giving a simple but broad overview.

\bibitem{harris64a} J. L. Harris, ``Resolving power and decision theory," J. Opt. Soc. Am. \textbf{54}, 606--611 (1964).

\bibitem{small09} An updated treatment of the one-point-source-or-two decision problem is given by Alexander R. Small, ``Theoretical limits on errors and acquisition rates in localizing switchable fluorophores," Biophys. J. \textbf{96}, L16--L18 (2008).

\bibitem{prasad14}  For a more formal Bayesian treatment, see S. Prasad, ``Asymptotics of Bayesian error probability and source super-localization in three dimensions," Opt. Expr. \textbf{22}, 16008--16028 (2014).

\bibitem{hell10} Stefan W. Hell, ``Far-field optical nanoscopy," Springer Series in Chemical Physics \textbf{96}, 365--398 (2010).

\bibitem{huang10} Bo Huang, Hazen Babcock, and Xiaowei Zhuang, ``Breaking the diffraction barrier: superresolution imaging of cells," Cell \textbf{143}, 1047--1058 (2010).

\bibitem{cremer13} Christoph Cremer and Barry R. Masters, ``Resolution enhancement techniques in microscopy," Eur. Phys. J. H \textbf{38}, 281--344 (2013).

\bibitem{hecht02} Eugene Hecht, \textit{Optics}, 4th ed. (Addison-Wesley, 2002), Ch.~13.

\bibitem{brooker02} Geoffrey Brooker, \textit{Modern Classical Optics} (Oxford Univ. Press, 2002), Ch.~12.

\bibitem{lipson11}  Ariel Lipson, Stephen G. Lipson, and Henry Lipson, \textit{Optical Physics}, 4th ed. (Cambridge Univ. Press, 2011), Ch.~12.  This edition of a well-established text adds a section on superresolution techniques, with a view that complements the one presented here.

\bibitem{axial}  Equation~\eqref{eq:abbe-limit} gives the \textit{lateral} resolution.  The resolution along the optical axis is poorer:  $d = \frac{\lambda}{n \, \sin^2 \alpha}$.

\bibitem{note:magnification}  However, the magnification of an objective does \underline{not} determine its resolution.

\bibitem{duffieux83} P. M. Duffieux, \textit{The Fourier Transform and Its Applications to Optics}, 2nd ed. (John Wiley \& Sons, 1983).  The first edition, in French, was published in 1946.  Duffieux formulated the idea of the optical transfer function in the 1930s.

\bibitem{lindberg12} Jari Lindberg, ``Mathematical concepts of optical superresolution," J. Opt. \textbf{14}, 083001 (23pp) (2012).

\bibitem{low-pass}  A subtle point:  the modulation transfer function is zero beyond a finite spatial frequency; yet the response in Fig.~\ref{fig:bode} is non-zero at all frequencies.  The explanation is that an object of finite extent has a Fraunhofer diffraction pattern (Fourier transform) that is \textit{analytic}, neglecting noise.  Analytic functions are determined by any finite interval (analytic continuation), meaning that one can, in principle, extrapolate the bandwidth and deduce the exact behavior beyond the cutoff from that inside the cutoff.  In practice, noise cuts off the information (Fig.~\ref{fig:bode}).  See Lucy\cite{lucy92} for a brief discussion and Goodman's book\cite{goodman05} for more detail.

\bibitem{lucy92} L. B. Lucy, ``Statistical limits to superresolution," Astron. Astrophys. \textbf{261}, 706--710 (1992).  Lucy does not assume the PSF width to be known and thus reaches the more pessimistic conclusion that $\Delta x \sim N^{-1/8}$. Since the second moments are then matched, one has to use the variance of the fourth moment to distinguish the images.

\bibitem{piche12} Kevin Pich\'e, Jonathan Leach, Allan S. Johnson, Jeff. Z. Salvail, Mikhail I. Kolobov, and Robert W. Boyd, ``Experimental realization of optical eigenmode superresolution," Opt. Exp. \textbf{20}, 26424 (2012).  Instruments with finite aperture sizes have discrete eigenmodes (that are not simple sines and cosines), which should be used for more accurate image restoration.

\bibitem{ramsay08} E. Ramsay, K. A. Serrels, A. J. Waddie, M. R. Taghizadeh, and D. T. Reid, ``Optical superresolution with aperture-function engineering," Am. J. Phys. \textbf{76}, 1002--1006 (2008).

\bibitem{toraldo55} G. Toraldo di Francia, ``Resolving power and information," J. Opt. Soc. Am. \textbf{45}, 497--501 (1955).

\bibitem{lipson03} S. G. Lipson, ``Why is superresolution so inefficient?" Micron \textbf{34}, 309--312 (2003).

\bibitem{lukosz66} W. Lukosz, ``Optical systems with resolving powers exceeding the classical limit," J. Opt. Soc. Am. \textbf{56}, 1463--1472 (1966).

\bibitem{lukosz67} W. Lukosz, ``Optical systems with resolving powers exceeding the classical limit.  II" J. Opt. Soc. Am. \textbf{57}, 932--941 (1967).

\bibitem{mukamel12} Eran A. Mukamel and Mark J. Schnitzer, ``Unified resolution bounds for conventional and stochastic localization fluorescence microscopy," Phys. Rev. Lett.~\textbf{109}, 168102 (2012).

\bibitem{fitzgerald12} James E. Fitzgerald, Ju Lu, and Mark J. Schnitzer, ``Estimation theoretic measure of resolution for stochastic localization microscopy," Phys. Rev. Lett.~\textbf{109}, 048102 (2012).

\bibitem{nieuwenhuizen13} Robert P. J. Nieuwenhuizen, Keith A. Lidke, Mark Bates, Daniela Leyton Puig, David Gr\"onwald, Sjoerd Stallinga, and Bernd Rieger, ``Measuring image resolution in optical nanoscopy," Nat. Meth. \textbf{10}, 557--562 (2013).

\bibitem{sivia06}  D. S. Sivia and J. Skilling, \textit{Data Analysis:  A Bayesian Tutorial}, 2nd ed. (Oxford Univ. Press, 2006), Ch.~5.

\bibitem{bobroff86} Norman Bobroff, ``Position measurement with a resolution and noise-limited instrument," Rev. Sci. Instrum. \textbf{57}, 1152--1157 (1986).

\bibitem{ober04} Raimund J. Ober, Sripad Ram, and E. Sally Ward, ``Localization accuracy in single-molecule microscopy," Biophys. J. \textbf{86}, 1185--1200 (2004).

\bibitem{mortensen10} Kim I. Mortensen, L. Stirling Churchman, James A. Spudich, and Henrik Flyvbjerg, ``Optimized localization analysis for single-molecule tracking and superresolution microscopy," Nat. Meth. \textbf{7}, 377--381 (2010).  Gives a useful assessment of various position estimators.

\bibitem{deschout14} Hendrik Deschout, Francesca Cella Zanacchi, Michael Mlodzianoski, Alberto Diaspro, Joerg Bewersdorf, Samuel T. Hess, and Kevin Braeckmans, ``Precisely and accurately localizing single emitters in fluorescence microscopy," Nat. Meth. \textbf{11}, 253--266 (2014).

\bibitem{yildiz05} Ahmet Yildiz and Paul R. Selvin, ``Fluorescence Imaging with One
Nanometer Accuracy:  Application to Molecular Motors," Acc. Chem. Res. \textbf{38}, 574--582 (2005).

\bibitem{patterson10} George Patterson, Michael Davidson, Suliana Manley, and Jennifer Lippincott-Schwartz, ``Superresolution imaging using single-molecule localization," Annu. Rev. Phys. Chem. \textbf{61}, 345--367 (2010).

\bibitem{errorbars}  One should set the errors to be the square root of the smooth distribution value deduced from the initial fit and then iterate the fitting process;\cite{norrelykke10} however, the conclusions would not change, in this case.

\bibitem{norrelykke10} Simon F. N{\o}rrelykke and Henrik Flyvbjerg, ``Power spectrum analysis with least-squares fitting:  {A}mplitude bias and its elimination, with application to optical tweezers and atomic force microscope cantilevers," Rev. Sci. Instrum. \textbf{81}, 075103 (2010).

\bibitem{betzig93} Eric Betzig and Robert J. Chichester, ``Single molecules observed by near-field scanning optical microscopy," Science \textbf{262}, 1422--1425 (1993).

\bibitem{ha99} Taekjip Ha, Ted. A. Laurence, Daniel S. Chemla, and Shimon Weiss, ``Polarization spectroscopy of single fluorescent molecules," J. Phys. Chem. B \textbf{103}, 6839--6850 (1999).

\bibitem{engelhardt11} Johann Engelhardt, Jan Keller, Patrick Hoyer, Matthias Reuss, Thorsten Staudt, and Stefan W. Hell, ``Molecular orientation affects localization accuracy in superresolution far-field fluorescence microscopy," Nano Lett. \textbf{11}, 209--213 (2011).

\bibitem{frantsuzov08} Pavel Frantsuzov, Masaru Kuno, Boldizs\'ar Jank\'o, and Rudolph A. Marcus, ``Universal emission intermittency in quantum dots, nanorods and nanowires," Nature  Phys. \textbf{4}, 519--522 (2008).


\bibitem{betzig06} Eric Betzig, George H. Patterson, Rachid Sougrat, O. Wolf Lindwasser, Scott Olenych, Juan S. Bonifacino, Michael W. Davidson, Jennifer Lippincott-Schwartz, and Harald F. Hess, ``Imaging intracellular fluorescent proteins at nanometer resolution," Science \textbf{313}, 1642--1645 (2006).

\bibitem{rust06} Michael J. Rust, Mark Bates, and Xiaowei Zhuang, ``Sub-diffraction-limit imaging by stochastic optical reconstruction microscopy (STORM)," Nat. Meth. \textbf{3}, 793--795 (2006).

\bibitem{precursors-stochastic}  Important precursors in using sequential localization to develop  stochastic localization techniques such as PALM and STORM were Qu et al.\cite{qu04} and Lidke et al.\cite{lidke05}  Stochastic localization was also independently developed by Hess et al.\cite{hess06}

\bibitem{qu04} Xiaohui Qu, David Wu, Laurens Mets, and Norbert F. Scherer, ``Nanometer-localized multiple single-molecule fluorescence microscopy," PNAS \textbf{101}, 11298--11303 (2004).

\bibitem{lidke05} Keith A. Lidke, Bernd Rieger, Thomas M. Jovin, and Rainer Heintzmann, ``Superresolution by localization of quantum dots using blinking statistics," Opt. Exp. \textbf{13}, 7052--7062 (2005).

\bibitem{hess06} Samuel T. Hess, Thanu P. K. Girirajan, and Michael D. Mason, ``Ultra-high resolution imaging by fluorescence photoactivation localization microscopy," Biophys. J. \textbf{91}, 4258--4272 (2006).

\bibitem{sparse} \textit{Sparseness} can improve resolution in other ways, as well.  For example, the new field of \textit{compressive sensing} also uses \textit{a priori} knowledge that a sparse representation exists in a clever way to improve resolution.\cite{babcock13}

\bibitem{babcock13} Hazen P. Babcock, Jeffrey  R. Moffitt, Yunlong Cao, and Xiaowei Zhuang, ``Fast compressed sensing analysis for super-resolution imaging using L1-homotopy," Opt. Expr. \textbf{21}, 28583--28596 (2013).

\bibitem{dertinger09}  T. Dertinger, R. Colyer, G. Iyer, S. Weiss, and J. Enderlein, ``Fast, background-free, 3D super-resolution optical fluctuation imaging (SOFI)," PNAS \textbf{106}, 22287--22292 (2009).  This clever technique uses intensity fluctuations due to multiple switching between two states of different brightness.

\bibitem{berro12}  Adam J. Berro, Andrew J. Berglund, Peter T. Carmichael, Jong Seung Kim, and J. Alexander Liddle, ``Super-resolution optical measurement of nanoscale photoacid distribution in lithographic materials," ACS Nano \textbf{6}, 9496--9502 (2012).  If one knows that vertical stripes are present, one can sum localizations by column to get a higher-resolution horizontal cross-section.

\bibitem{hell94} Stefan W. Hell and Jan Wichmann, ``Breaking the diffraction resolution limit by stimulated emission: stimulated-emission-depletion fluorescence microscopy," Opt. Lett. \textbf{19}, 780--782 (1994).

\bibitem{harke08} Benjamin Harke, Jan Keller, Chaitanya K. Ullal, Volker Westphal, Andreas Sch\"onle, and Stefan W. Hell, ``Resolution scaling in STED microscopy," Opt. Expr. \textbf{16}, 4154--4162 (2008).

\bibitem{dyba05} M. Dyba, J. Keller, and S. W. Hell, ``Phase filter enhanced STED-4Pi fluoroescence microscopy:  theory and experiment," New J. Phys. \textbf{7}, 134 (2005).

\bibitem{pawley06} J.B. Pawley, \textit{Handbook of Biological Confocal Microscopy}, 2nd ed. (Springer, 2006).

\bibitem{diaspro05} Alberto Diaspro, Giuseppe Chirico, and Maddalena Collini, ``Two-photon fluorescence excitation and related techniques in biological microscopy," Quart. Rev. Biophys. \textbf{38}, 97--166 (2005).

\bibitem{cremer78} C. Cremer and T. Cremer, ``Considerations on a laser-scanning-microscope with high resolution and depth of field," Microsc. Acta \textbf{81}, 31--44 (1978).

\bibitem{hell92} Stefan Hell, Ernst H.K. Stelzer, ``Fundamental improvement of resolution with a 4Pi-confocal fluorescence microscope using two-photon excitation," Opt. Comm. \textbf{93}, 277--282 (1992).

\bibitem{synge28} E. H. Synge, ``A suggested method for extending microscopic resolution into the ultra-microscopic region," Philos. Mag. \textbf{6}, 356--362 (1928).

\bibitem{betzig86} E. Betzig, A. Lewis, A. Harootunian, M. Isaacson, and E. Kratschmer, ``Near-field scanning optical microscopy (NSOM): development and biophysical applications," Biophys. J. \textbf{49}, 269--279 (1986).

\bibitem{novotny12} Lukas Novotny and Bert Hecht, \textit{Principles of Nano-Optics}, 2nd ed. (Cambridge Univ. Press, 2012).

\bibitem{simonetti06} F. Simonetti, ``Multiple scattering: The key to unravel the subwavelength world from the far-field pattern of a scattered wave," Phys. Rev. E \textbf{73}, 036619 (2006).

\bibitem{heintzmann02} Rainer Heintzmann and Thomas M. Jovin, and Christoph Cremer, ``Saturated patterned excitation microscopy---a concept for optical resolution improvement," J. Opt. Soc. Am. A \textbf{19}, 1599--1609 (2002).

\bibitem{fujita07} Katsumasa Fujita, Minoru Kobayashi, Shogo Kawano, Masahito Yamanaka, and Satoshi Kawata, ``High-resolution confocal microscopy by saturated excitation of fluorescence," Phys. Rev. Lett. \textbf{99}, 228105 (2007).

\bibitem{pendry00} J. B. Pendry, ``Negative refraction makes a perfect lens," Phys. Rev. Lett. \textbf{85}, 3966-3969 (2000).

\bibitem{fang05} Nicholas Fang, Hyesog Lee, Cheng Sun, Xiang Zhang, ``SubÐdiffraction-limited optical imaging with a silver superlens," Science \textbf{308}, 534--537 (2005).

\bibitem{gustafsson00} M. G. L. Gustafsson, ``Surpassing the lateral resolution limit by a factor of two using structured illumination microscopy," J. Microscopy \textbf{198}, 82--87 (2000).

\bibitem{gustafsson05} Mats G. L. Gustafsson, ``Nonlinear structured-illumination microscopy:  Wide-field fluorescence imaging with theoretically unlimited resolution," Proc.Nat. Acad. Sci. (USA) \textbf{102}, 13081--13086 (2005).

\bibitem{boto00} Agedi N. Boto, Pieter Kok, Daniel S. Abrams, Samuel L. Braunstein, Colin P. Williams, and Jonathan P. Dowling, ``Quantum interferometric optical lithography:  exploiting entanglement to beat the diffraction limit," Phys. Rev. Lett. \textbf{85}, 2733--2736 (2000).

\bibitem{rozema14} Lee A. Rozema, James D. Bateman, Dylan H. Mahler, Ryo Okamoto, Amir Feizpour, Alex Hayat, and Aephraim M. Steinberg, ``Scalable spatial superresolution using entangled photons," Phys. Rev. Lett. \textbf{112}, 223602 (2014).

\bibitem{oppenheim13}  Jacob N. Oppenheim and Marcelo O. Magnasco, ``Human time-frequency acuity beats the {F}ourier {U}ncertainty principle," Phys. Rev. Lett. \textbf{110}, 044301 (2013).

\end{thebibliography}
\end{document}